# MOBILITY21: Strategic Investments for Transportation Infrastructure & Technology

| Rahul Mangharam | Megan Reyerson | Steve Viscelli |
| University of Pennsylvania | University of Pennsylvania | University of Pennsylvania |

| Hamsa Balakrishanan | Alexandre Bayen | Surabh Amin |
| MIT | UC Berkeley, ITS | MIT |

| Leslie Richards | Leo Bagley | George Pappas |
| PennDOT | PennDOT | University of Pennsylvania |

America's transportation infrastructure is the backbone of our economy. A strong infrastructure means a strong America – an America that competes globally, supports local and regional economic development, and creates jobs. Strategic investments in our transportation infrastructure are vital to our national security, economic growth, transportation safety and our technology leadership.[1] This document outlines critical needs for our transportation infrastructure, identifies new technology drivers and proposes strategic investments for safe and efficient air, ground, rail and marine mobility of people and goods. Three axes drive fundamental changes into how we will plan, deploy and use our transportation networks through the century:

**1. New Vehicle Platforms**: With the rapid development of Connected/Automated Vehicles, Delivery Drones and Commercial Space we identify the need for (i) a smart rebuild of the interstate system, (ii) an integrated transportation communications infrastructure and (iii) a nation-wide network of autonomous truck ports to substantially reduce the need to build expensive new roads. This is infrastructure that can be built faster, encourages innovation and unleashes the next revolution in mobility.

**2. Infrastructure Integrated with Internet-of-Things**: Widespread sensor deployment within transportation networks will enable new services such as smart parking, matching drivers with spots, and congestion pricing, matching road capacity to trip demand to achieve shorter, more predictable travel. Real-time data on use and changing transportation needs will provide scalable analytics for making existing transportation networks resilient to aging and natural disasters. The use of such sensors, real-time data and Artificial Intelligence in our vehicles and transportation networks requires policy and infrastructure for (i) ensuring autonomous systems make safe, fair and predictable decisions, (ii) efficient exchange of travel data while ensuring user privacy and asset security, and (iii) co-existence of private and public autonomous systems with an integrated technology roadmap.

**3. New Transportation Users**: New shared and flexible mobility models such as "Mobility as a Service" and "When-I-Want-It/Where-I-Want-It Logistics" require technology platforms (i) for infrastructure service partnerships in air, ground, marine and rail, and (ii) technology services for wider adoption of "sharing economy" transportation modes to leverage underutilized assets and put them to productive use.

As transportation technology evolves, we propose an "Integrated Deployment" approach for performance-based regulations rather than the current prescriptive policy and planning. This approach includes the clustering of innovations within technology ecosystems, and in the case of many infrastructure-based technologies, through public-private partnerships. Such integrated deployments represent investments where the productivity gains produced are greater than the sum of their individual technology parts.

---

[1] Mynatt et al. (2017) "A National Research Agenda for Intelligent Infrastructure" CCC Led Whitepapers
http://cra.org/ccc/resources/ccc-led-whitepapers/, last accessed April 12, 2017.

A.1. Ground Transportation: **Safe Autonomous Platforms**
As the rapid introduction of partially autonomous "Autopilot" Advanced Driver Assistance Systems (ADAS) are already outpacing current safety certification and regulatory standards, there is a need to build public confidence prior to the mainstream introduction of autonomous vehicles. The Transportation and Infrastructure Committee must establish a technology and policy foundation for the deployment of automated vehicles by achieving broad industry participation in the NHTSA Automated Vehicles Policy development process. This approach will require engaging with States, safety regulators, and advocates in establishing processes that would help the transition from testing to larger scale deployment. The development of a clear technology and policy roadmap in the Federal Motor Vehicle Safety Standards (e.g. for Automated Vehicle, ADAS, visibility, lighting, etc.) will fast track innovations such as the development of machine learning algorithms which make safe, fair and predictable decisions, scalable verification of software and software updates in AVs, and an insurance framework for AVs.

A.2. Ground Transportation: **Smart Rebuild of the Interstate System**
The states and the nation face a 50+ year old 46,000-mile interstate system in need of major reconstruction and rehabilitation but without the adequate funding necessary to plan and execute a comprehensive, multi-year, multi-state program. The Smart Rebuild of the Interstate System should bring the system up to modern highway and bridge design standards, include critical operational and safety improvements, plan strategic widenings and advanced traffic management where appropriate geared toward eliminating or reducing major commuter and freight bottlenecks, and include connected and autonomous vehicle infrastructure necessary for the next generation of vehicles.

A.3. Ground Transportation: **Communications Infrastructures for Smart Cities**
Smart City developments are attractive as they enhance the quality of life in ultra-dense urban areas, create jobs and facilitate high throughput economic activity with rural regions. A foundation for such 21$^{st}$ Century data-driven Smart Cities is to have an integrated Broadband Infrastructure and Secure Shared Spectrum to support connected vehicles and flexible logistics. It is essential to include broadband 5G and wide area low-rate (LoRa) networks in any infrastructure legislation, including broadband funding for rural or otherwise hard-to-serve areas and support for technology-driven approaches to spectrum sharing between fixed and vehicular networks for coordinated safety, improved utilization and adaptation to disruptive technologies.

A.4. Ground Transportation: **Freight Infrastructure Investment**
Autonomous trucks are poised to revolutionize the work of the freight industry, resulting in dramatic productivity gains and fundamental changes to trucking operations. This unprecedented opportunity for public investments in infrastructure offers foster faster, safer and cleaner trucking through the more efficient use of existing road capacity. The U.S. should build a network of Urban Truck Ports at key points of road congestion around our major cities. Like a port where the ocean meets the land, truck ports would be placed where the urban meets the rural. Ports would be a place where different tractors—one designed for rural areas and one for urban areas—can meet up to swap trailers. By splitting up trips by the type of driving required and the best tractor for each, these ports would allow the industry to invest in the best technology for each kind of driving, avoid driving through rush hour, and bring drivers home more often. Truck ports could save the industry billions of dollars annually in fuel, congestion, and driver turnover costs. And by utilizing our most congested roads at off-peak times, a nation-wide network of ports would substantially reduce the need to build new roads.

A.5. Ground Transportation: **Addressing Security & Privacy Gaps which put American Drivers at Risk**
Today's vehicle functionality, safety, and privacy all depend on the functions of more than 50 separate electronic control units within the vehicle, as well as communication between vehicles for active networked safety. As vehicles continue to capture and transmit driving data across vehicular networks, there are more avenues through which a hacker could introduce malicious code, and through which a driver's basic right to privacy could be compromised. Currently, most automobile manufacturers were unaware of, or unable to report on past hacking incidents. Security measures to prevent remote access to vehicle electronics are inconsistent and haphazard across all automobile manufacturers. The Transportation and Infrastructure Committee must guide robust vehicle security policies to ensure every vehicle manufacturer adopt a common set of security and privacy principles which include (1) transparency, (2) choice, (3) respect for context, (4) data minimization, de-identification and retention, (5) data security, (6) integrity and access, and (7) accountability.

B.1. Rail Transportation: **Intercity Passenger and Freight Rail Service**
The nation needs a complement to our interstate system to carry passengers and freight. Investments in improving passenger rail, primarily in the Amtrak Northeast Corridor where significant planning has occurred, but also necessary to support the efforts of some states to improve Amtrak service or establish high speed rail. Rail freight improvements, in partnership with Amtrak and the major railroads and seaports, would benefit the new on-demand national and international economy. Rail freight bottlenecks can affect not only freight movements but passenger traffic and should be mitigated as part of strategic transportation investments.

B.2. Rail Transportation: **Integrated Technology Plan**
The Rail Industry is undergoing a silent revolution with system-wide sensors for real-time equipment monitoring and the use of drones for inspection and crash analysis of rail assets for hazard monitoring. Sensor technologies such as wheel encoders for early warning of railcar maintenance and wayside visual and ultrasound detectors for detecting rail flaws enable the same assets to last longer. The Transportation and Infrastructure Committee must focus efforts on how to stretch current infrastructure by the widespread use of Electronically Controlled Pneumatic Braking for coordinated braking across multiple cars, remote inspection of brake health, and for graduated release in challenging weather conditions. A second coordinated effort is needed on Rail System Data Analytics to enhance safety and throughput with rail-to-rail trip optimizers, network level movement optimizers, monitoring the health of rolling stock, and with the use of uniquely identified freight cars and locomotives.

C. Air Transportation
In 2013, the FAA reported that the 485 commercial airports in the U.S. support 9.6 million jobs; create an annual payroll of $358 billion; and produce an annual output of $1.1 trillion. However, the cost of delay due to supply and demand imbalances and the lack of air traffic control efficiency cost the entire system $32.9 billion (in 2007), over half of which was borne by U.S. air passengers ($16.7 billion). As 42% of aircraft delays are due to maintenance, supply chain, operations and ground handling issues and 25% come from airport and airspace congestion, the benefits of improved infrastructure and connectivity across components of the airport infrastructure are clear.

C.1. Air Transportation: **Improving aviation infrastructure**
The system of U.S. airports does not universally struggle from a lack of infrastructure, but rather, from a mismatch between supply and demand. The nation's biggest airports, our international hubs, struggle with surging demand and inefficient ground and landside infrastructure; these airports clearly need investment in terms of new hard infrastructure, particularly as other countries look to build up their own international airport hubs. Yet, recent findings show that small cities within 300 miles of a major hub airport are unable to build their own air service if their somewhat proximate hub airport grows service; travelers find it more attractive to drive to the hub airport and take advantage of higher levels of service. We must make investments in airport infrastructure with the understanding that airports are nodes in a larger, interconnected multimodal transportation system; moreover, that investments in one airport may be detrimental to the air service and economic development of a neighboring region. In considering modernization efforts, a careful analysis of the question of whether or not (and to what extent) to privatize ATC functions, the associated risks and benefits, and the technological approach is much needed.

C.2. Air Transportation: **Unmanned Aerial Systems**
Unmanned aerial systems also hold the potential to revolutionize urban and rural parcel delivery; medical device testing and drug delivery; the agricultural sector through precision farming; and the oil and gas industry through efficient pipeline inspections and infrastructure maintenance. While the FAA has not yet approved the certification of such UAS vehicles for commercial operations, both the public and private sectors are hard at work developing certifications and UAS technologies. The integration of UAS into the nation's airspace system has been estimated to have an economic benefit of more than $13.6 billion in the first three years alone, in addition to creating 34,000 manufacturing jobs and 70,000 new jobs. There must be research and investment in the air traffic control infrastructures – hard infrastructures such as landing locations and new distribution centers, as well as communications, navigation and surveillance infrastructure – so that cities can ensure their competitiveness in the future.

**Actions and Recommendations**
Based on the preceding discussion, we offer the following recommended actions:
1. Analyze skills and education requirements to facilitate new technical jobs for shared, autonomous and data-driven technologies in road and air technology management and maintenance due to a looming talent and skills shortage in the planning, design and construction of transportation technologies.

2. Regulate safe autonomous systems, Vehicle-to-Vehicle (V2V) and Vehicle-to-Infrastructure (V2I) security and privacy in a way that aligns with National Highway Traffic Safety Administration (NHTSA).

3. Expand freight mobility by establishing Unmanned Systems, such as Ground Delivery "Bots" and Unmanned Aerial Vehicles (UAVs), as a new freight mode. Address constraints to UAVs in next Federal aviation reauthorization and encourage testing and safety/traffic management models for industry and public sector cooperation. For ground delivery unmanned robots, encourage a "complete streets" design approach to their integration.

4. Bridge gaps between research & development and deployment (R&D&D) of shared, autonomous and data-driven technologies: Advocate for more collaborative investment in high-risk/high-reward R&D and operational testing. Use "challenges" and other competitions to foster innovation in getting new technologies and integrated systems deployed. Establish a clear roadmap for architecture and standards that are needed to accelerate technology deployment from a commercial/public procurement perspective. Identify needs to create or maintain architecture and standards that assure quality, safety, security, accessibility, interoperability, and reliability of products, processes, and services for critical initiatives—such as connected and automated vehicles, or systems that address vulnerable road users (e.g. pedestrians, cyclists, etc.).

5. Increase overall investments in transportation with a greater focus on technology: Support new and long-term sustainable funding and financing in transportation infrastructure that expand technology-driven mobility investments and preserves the broad ITS eligibilities under the Fixing America's Surface Transportation Act (FAST Act) to fund technology capital projects and operations and maintenance.

6. Leverage technology to address the problem of Distracted Driving with practical solutions, which work on changing driving culture. Encourage more research to understand the complexity of Distracted Driving and the factors that elevate crash risk (or potentially reduce crash risk as in the case of safety technologies). Where necessary, establish industry guidelines for new types of driver interfaces, such as voice command systems.

*This material is based upon work supported by the National Science Foundation under Grant No. 1136993. Any opinions, findings, and conclusions or recommendations expressed in this material are those of the authors and do not necessarily reflect the views of the National Science Foundation.*